\documentclass[onecolumn,accepted=2021-06-16]{quantumarticle}
\usepackage{amsmath}
\usepackage{amssymb}
\usepackage{amsthm}
\usepackage{amsfonts}
\usepackage[caption=false]{subfig}
\usepackage[colorlinks]{hyperref}
\usepackage[all]{hypcap}
\usepackage{tikz}
\usepackage{relsize}
\usepackage{color,soul}
\usepackage[utf8]{inputenc}
\usepackage{capt-of}
\usepackage[numbers]{natbib}
\usepackage{float}
\usepackage{listings}
\usepackage[T1]{fontenc}  % Without this, underscores lost when copy-pasting output!
\usetikzlibrary{decorations.pathreplacing}

\DeclareFixedFont{\ttb}{T1}{txtt}{bx}{n}{10}
\DeclareFixedFont{\ttm}{T1}{txtt}{m}{n}{10}
\definecolor{deepblue}{rgb}{0,0,0.5}
\definecolor{deepred}{rgb}{0.6,0,0}
\definecolor{deepgreen}{rgb}{0,0.5,0}
\newcommand\cppstyle{\lstset{
language=C++,
basicstyle=\ttm,
otherkeywords={uint8_t, __m256i, size_t, ASSERT_TRUE, EXPECT_TRUE, TEST, BENCHMARK},
keywordstyle=\ttb\color{deepblue},
emphstyle=\ttb\color{deepblue},
stringstyle=\color{deepgreen},
commentstyle=\fontfamily{txtt}\selectfont\color{gray},
showstringspaces=false,
literate={*}{{\char42}}1
         {-}{{\char45}}1
}}
\lstnewenvironment{cpp}[1][]
{\cppstyle\lstset{#1}}{}
\newcommand\pythonstyle{\lstset{
language=python,
basicstyle=\ttm,
morekeywords={assert,as,echo},
keywordstyle=\ttb\color{deepblue},
emphstyle=\ttb\color{deepblue},
stringstyle=\color{deepgreen},
commentstyle=\fontfamily{txtt}\selectfont\color{gray},
showstringspaces=false,
literate={*}{{\char42}}1
         {-}{{\char45}}1
}}
\lstnewenvironment{python}[1][]
{\pythonstyle\lstset{#1}}{}

\renewcommand\thesection{\arabic{section}}

\theoremstyle{definition}

\theoremstyle{definition}

\theoremstyle{definition}

\newcommand{\eq}[1]{\hyperref[eq:#1]{Equation~\ref*{eq:#1}}}
\renewcommand{\sec}[1]{\hyperref[sec:#1]{Section~\ref*{sec:#1}}}
\DeclareRobustCommand{\app}[1]{\hyperref[app:#1]{Appendix~\ref*{app:#1}}}
\newcommand{\fig}[1]{\hyperref[fig:#1]{Figure~\ref*{fig:#1}}}
\newcommand{\tbl}[1]{\hyperref[tbl:#1]{Table~\ref*{tbl:#1}}}
\newcommand{\theoremref}[1]{\hyperref[theorem:#1]{Theorem~\ref*{theorem:#1}}}
\newcommand{\definitionref}[1]{\hyperref[definition:#1]{Definition~\ref*{definition:#1}}}

\begin{document}
\title{Stim: a fast stabilizer circuit simulator}

\date{\today}
\author{Craig Gidney}
\email{craiggidney@google.com}
\affiliation{Google Inc., Santa Barbara, California 93117, USA}

\begin{abstract}
This paper presents ``Stim", a fast simulator for quantum stabilizer circuits.
The paper explains how Stim works and compares it to existing tools.
With no foreknowledge, Stim can analyze a distance 100 surface code circuit (20 thousand qubits, 8 million gates, 1 million measurements) in 15 seconds and then begin sampling full circuit shots at a rate of 1 kHz.
Stim uses a stabilizer tableau representation, similar to Aaronson and Gottesman's CHP simulator, but with three main improvements.
First, Stim improves the asymptotic complexity of deterministic measurement from quadratic to linear by tracking the {\em inverse} of the circuit's stabilizer tableau.
Second, Stim improves the constant factors of the algorithm by using a cache-friendly data layout and 256 bit wide SIMD instructions.
Third, Stim only uses expensive stabilizer tableau simulation to create an initial reference sample.
Further samples are collected in bulk by using that sample as a reference for batches of Pauli frames propagating through the circuit.
\end{abstract}

\maketitle

\emph{
Readers who want to try Stim can find its source code in this paper's ancillary files or on github at \url{https://github.com/quantumlib/stim}.
Stim is also available as a Python 3 pypi package installed via ``}\texttt{pip install stim}\emph{".}
\emph{There is also a pypi package ``stimcirq" that exposes stim as a cirq sampler}.

\section{Introduction}
\label{sec:introduction}

In the field of human computer interaction, an important metric is the delay between human action and computer reaction.
Quoting Jakob Nielsen \cite{nielsen1994usability,nielsenusabilitysite1993}:

\begin{itemize}
    \item 0.1 seconds is about the limit for having the user feel that the system is reacting instantaneously [...]
    \item 1.0 second is about the limit for the user's flow of thought to stay uninterrupted [...]
    \item 10 seconds is about the limit for keeping the user's attention [...]
\end{itemize}

I bring this up because the underlying motivation driving this paper is that I want to tinker with huge stabilizer circuits, and I want to stand in the 0.1s bucket while I do it.
For example, consider the single level 15-to-1 surface code T state factory shown in figure 4 of \cite{gidney2019catalyzeddistillation}.
It corresponds to a circuit with millions of gates spanning 15 thousand qubits.
What would it take to simulate {\em that} in 0.1 seconds, with no built-in preconceptions about the surface code, while a user makes iterative changes that break and restore the functionality of the circuit?
There are several obstacles that make this difficult.

\begin{figure}
    \centering
    \resizebox{\linewidth}{!}{
        \includegraphics{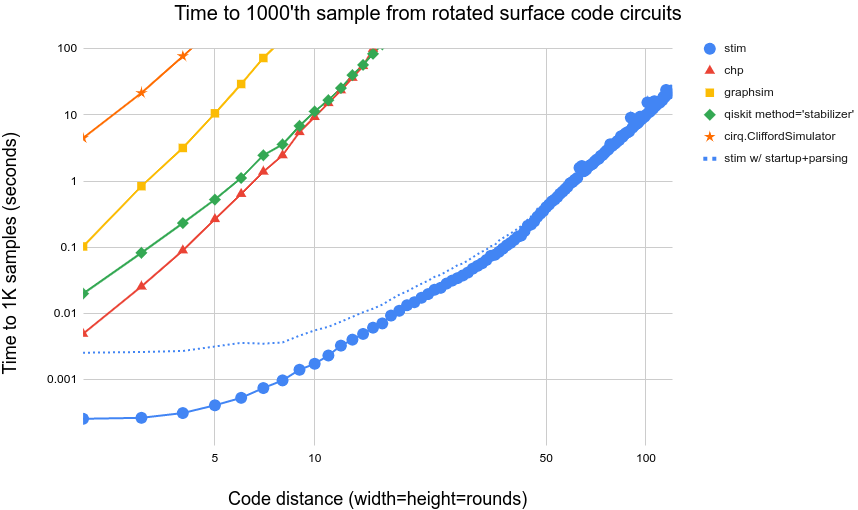}
    }
    \caption{
        Delay until one thousandth sample on $d \times d \times d$ surface code memory circuits \cite{horsman2012latticesurgery}.
        The intention of this benchmark is to highlight simulators that can perform some sort of compilation or analysis that lowers the cost of taking multiple samples.
        Stim does particularly well on this benchmark, with timings almost identical to its time-to-first-sample.
        This is because it uses its first sample as a reference for batches of hundreds or thousands of Pauli frame simulations that are updated in parallel using SIMD instructions.
        The per-sample performance of Qiskit's simulator also improves when collecting more than one sample.
    }
    \label{fig:bench-surface-1000}
\end{figure}

First, if a circuit has a million gates, and your computer's clock speed is 2 gigahertz, then the simulator has a budget of 200 clock cycles per gate.
That's a tight budget.

Second, describing an arbitrary $n$-qubit stabilizer state takes at least $\frac{1}{2} n^2$ bits \cite{gross2006hudson,karanjai2018contextuality,howmanystabilizers2021}.
This means that, in the worst case, a general stabilizer simulator that works in an online fashion will need megabytes of information to track the state of a 15000 qubit system.
Storing this amount of information is not a problem.
The problem is repeatedly touching that information without blowing the budget on cache misses.

Third, typically, simulating a Clifford gate has complexity $\Omega(n)$ where $n$ is the number of qubits in the circuit.
Given the aforementioned cycle budget, the constant factor hiding behind that $\Omega$ has to be roughly 0.01 cycles.

Fourth, the surface code is chock full of measurements and, basically, historically, the cost of measurements in stabilizer simulators started at $\Theta(n^3)$ \cite{gottesman1997stabilizerformalism}, improved to $\Theta(n^2)$~\cite{aaronson2004chp}, and then stayed there.
Aaronson and Gottesman's~\cite{aaronson2004chp} CHP simulator?
Measurement takes $\Theta(n^2)$ time in the worst case.
Anders and Briegel's~\cite{anders2006fastgraphsim} graph state simulator?
Measurement takes $\Theta(n^2)$ time in the worst case.
Bravyi et al's~\cite{bravyi2019simulation} CH form simulator?
Measurement takes $\Theta(n^2)$ time in the worst case.
We can't afford $\Theta(n^2)$ time.

Fifth, the circuit I've been talking about isn't even technically a stabilizer circuit.
It contains magic state injections.
This obstacle has actually been the subject of a lot of recent work \cite{bravyi2019simulation,bu2019efficient,huang2020feynman,huang2019approximate}.
I'll be completely ignoring it in this paper.

Sixth, when simulating an error correcting circuit, it's common to sample the circuit millions or even trillions of times.
For example, going to large code distances can allow a lone half distance error mechanism to shine through the background haze of full distance errors.
But logical errors become rare at large code distances, and so many samples have to be taken to see them.
Based on discussions I've had with co-workers who have done these sorts of large scale simulations \cite{conversationmikeasutin}, the main bottleneck is not decoding the errors but rather (surprisingly) generating the samples.
So it's not just the latency until the first sample that matters; the sustained bulk sampling rate is also important.

With all those obstacles listed I should note that, of course, by using hardcoded knowledge about magic state distillation and the surface code, it's possible to make a specialized simulator that could solve my example case much faster than a general stabilizer circuit simulator.
But my underlying motivation is one of exploration, and I would strongly prefer to create a tool whose performance and correctness doesn't require understanding the behavior of places I haven't been yet.
So, in this paper, I present my failed attempt to reach my ridiculous goal.
The result is a software tool I've named "Stim".
Although Stim can't simulate a 15 thousand qubit surface code circuit with millions of gates in under a tenth of a second, it can at least get that job done in roughly ten seconds.
Then, with that initial sample acquired, it can begin producing batches of hundreds of samples in a tenth of the time.

Stim makes three key improvements over previous stabilizer simulators.
These improvements address obstacle 4 (the measurement obstacle), obstacle 3 (the constant factor obstacle), and obstacle 6 (the mega sampling obstacle).

The measurement obstacle is addressed by making an algorithmic improvement that decreases the time complexity of {\em deterministic} measurements from quadratic to linear.
This is beneficial in contexts, such as the surface code, where a circuit has many measurements but those measurements are almost entirely deterministic.

The constant factor obstacle is addressed by using SIMD (Same Instruction Multiple Data) instructions in tight loops iterating over contiguous memory.
Using SIMD is not a novel concept (e.g. Aaronson et al's CHP simulator \cite{aaronson2004chp} packs the bits from Pauli terms into bytes, which allows 8 to be operated on by each instruction) but it's technically challenging to implement well and Stim does it at larger register sizes than previous work.
Specifically, Stim uses 256 bit wide Advanced Vector Extensions (AVX) \cite{wiki:Advanced_Vector_Extensions} and also I spent a lot of effort minimizing the number of operations in key loops.
For example, a core operation for most gates is multiplying together strings of Pauli terms, which requires accumulating scalar factors from individual Pauli multiplications.
I found a way to do this using only 11 bitwise operations per group of Paulis (see \fig{pauli_mult_code}).
Modern Intel CPUs can apply 3 bitwise AVX operations per clock cycle \cite{intel-intrinsics-and}, and 256 is much larger than 11, suggesting that a budget of 0.01 cycles per qubit during a Clifford gate isn't quite as ridiculous as it seems.

The mega sampling obstacle is addressed by including a Pauli frame simulator that uses SIMD instructions to operate on the state of hundreds of simulations in parallel.
This amortizes the cost of cache misses and branch mispredictions and other performance killers.
This is a simple enough idea, but it gives Stim an enormous advantage over previous work when bulk sampling (see \fig{bench-surface-1000}).

This paper is divided into sections.
In \sec{concepts}, I explain key background concepts that are needed in order to understand the simulation strategies being described.
Then in \sec{framesim} and \sec{tableausim} I describe the Pauli frame simulator and stabilizer tableau simulators implemented by Stim.
\sec{engineering} describes several of the software engineering details that went into Stim, such as why I decided to use a dense representation instead of a sparse representation and how I went about testing that operations were implemented correctly.
Stim's performance is compared to other stabilizer simulators in \sec{compare}.
Finally, \sec{conclusion} wraps everything up.
There are also a few short usage examples in \app{example}.

\section{Key Concepts}
\label{sec:concepts}

\subsection{Pauli Products}

There are four Pauli gates: the identity $I = \begin{bmatrix} 1&0\\0&1\end{bmatrix}$, the bit flip $X = \begin{bmatrix} 0&1\\1&0\end{bmatrix}$, the phase flip $Z = \begin{bmatrix} 1&0\\0&-1\end{bmatrix}$, and the other one $Y = \begin{bmatrix} 0&-i\\i&0\end{bmatrix}$.
Sometimes the identity gate is not considered to be a Pauli gate, but in this paper it is.

A Pauli product is a set of Pauli gates applied to qubits.
For example, $X_1 \cdot Y_2 \cdot Z_3$ is a Pauli product that applies $X$ to qubit 1, $Y$ to qubit 2, $Z$ to qubit 3, and $I$ to everything else.
The set of all Pauli products form a group under matrix multiplication.
This group is called the Pauli group.

Paulis can be encoded in a variety of ways.
One encoding, which I'll be calling the ``xz encoding", represents a Pauli by decomposing it into an $x$ bit representing the presence of an $X$ gate, a $z$ bit representing the presence of a $Z$ gate, and an ignored global phase:

$$\texttt{encode\_xz}(P) = (x, z) = (P \notin \{I, Z\}, P \notin \{I, X\}) = \texttt{encode\_xz}(X^x Z^z)$$

Meaning:

$$\texttt{encode\_xz}(I) = (0, 0)$$
$$\texttt{encode\_xz}(X) = (1, 0)$$
$$\texttt{encode\_xz}(Y) = (1, 1)$$
$$\texttt{encode\_xz}(Z) = (0, 1)$$

In the xz encoding (and most others), two Paulis can be multiplied up to global phase by xoring their encoded representations:

$$\texttt{encode\_xz}(P_1 P_2) = \texttt{encode\_xz}(P_1) \oplus \texttt{encode\_xz}(P_2) = (x_1 \oplus x_2, z_1 \oplus z_2)$$

In Stim, Pauli products are represented by the \path{stim::PauliString} class.

\subsection{Stabilizer Generators}

When a quantum state is in the +1 eigenstate of an operation, that operation is said to be a {\em stabilizer} of the state.
For example: $+Z$ is a stabilizer of the computational basis state $|0\rangle$, $+Z$ is {\em not} a stabilizer of the computational basis state $|1\rangle$, $+X$ is a stabilizer of the superposed state $|0\rangle + |1\rangle$, and $-YY$ is a stabilizer of the entangled state $|00\rangle + |11\rangle$.
Stabilizer operations that are Pauli products are of particular interest, as they form the basis of the {\em stabilizer formalism} \cite{gottesman1997stabilizerformalism} (easily one of the most important and foundational ideas in quantum error correction).
In this paper, whenever I say "stabilizer" I always mean "Pauli product stabilizer".

The stabilizers of a state form a group under multiplication.
Let $A$ and $B$ be stabilizers of a state $|\psi\rangle$, meaning $A |\psi\rangle = |\psi\rangle = B |\psi\rangle$.
$AB$ is also a stabilizer of $|\psi\rangle$, because $A B|\psi\rangle = A |\psi\rangle = |\psi\rangle$.

The stabilizers of a state commute.
Note that $A B|\psi\rangle = B A|\psi\rangle$, so $0 = (AB - BA) |\psi\rangle = [A, B] |\psi\rangle$.
The commutator of $A$ and $B$ must project $|\psi\rangle$ to the zero vector.
The commutators of Pauli products are always scalar factors of Pauli products, where only the scalar factor can affect the length of $|\psi\rangle$, so it must be the case that the scalar factor is zero, meaning $[A, B] = 0$ (i.e. $A$ and $B$ commute).

If you start with the set of all stabilizers of a state, and iteratively remove stabilizers that can be formed by multiplying together other stabilizers, you are left with a small set of stabilizers referred to as {\em generators} of the state's stabilizer group.
For example, $|000\rangle$ has seven stabilizers (not counting the vacuous identity stabilizer) and these stabilizers can be generated by $ZII$, $IZI$, and $IIZ$.
An $n$ qubit state will have at most $n$ stabilizer generators.
An $n$ qubit state with exactly $n$ stabilizer generators is a {\em stabilizer state}.

\subsection{Stabilizer Tableaus}

A {\em Clifford operation} is a unitary quantum operation that conjugates Pauli products into Pauli products.
$C$ is Clifford if, for all pauli products $P$, it is the case that $C^\dagger P C$ is also a Pauli product.
In fact, a Clifford operation can be uniquely identified (up to global phase) by how it conjugates Pauli products.

A stabilizer tableau is a representation of a Clifford operation that simply directly stores how the Clifford operation conjugates each generator of the Pauli group.
In this paper, I use the generators $X_q$ and $Z_q$ for each qubit $q$ that the operation touches.

For example, here is a stabilizer tableau for the Controlled Y gate $C_Y$:

$$
\text{tableau}(C_Y) = \begin{array}{r|cc|cc}
    & X_1 & Z_1 & X_2 & Z_2   \\
    \hline
    \pm & + & + & + & + \\
    1 & X & Z & Z & Z \\
    2 & Y &   & X & Z \\
\end{array}
$$

Each column describes how $C_Y$ conjugates one of the four generators of the two qubit Pauli group.
The column labelled $X_1$ states that $C_Y$ conjugates $X_1$ into $+X_1 Y_2$, i.e. that $C_Y^{-1} X_1 C_Y = X_1 Y_2$.

Any qubit not mentioned by a tableau is unaffected by that tableau.
For example, if $X_q$ does not appear in the tableau then it is understood that the tableau's operation conjugates $X_q$ into $X_q$.

In order for a stabilizer tableau to be valid, i.e. to represent a Clifford operation, it must preserve commutativity and anticommutativity.
The Pauli products in its columns must commute or anticommute in the same way that its generators do.
The column for $X_a$ must commute with the column for $X_b$ and the column for $Z_b$, but must anticommute with the column for $Z_a$.
Also, a valid tableau doesn't have missing columns.
If there is a row for qubit $q$, there must be columns for the generators $X_q$ and $Z_q$.

Storing an $n$-qubit stabilizer tableau uses $4n^2 + O(n)$ bits.
There are $2n$ generator outputs to store, each output has $n$ Pauli terms, and each Pauli term is xz-encoded into two bits.

In Stim, stabilizer tableaus are represented by the class \path{stim::Tableau}.
Storing the stabilizer tableau is Stim's dominant space cost.
All other space costs are linear in the number of qubits.

\subsubsection{Conjugating a Pauli Product by a Tableau}

To conjugate a Pauli product observable by the Clifford operation represented by a stabilizer tableau, start by decomposing the Pauli product into the generators listed in the tableau.
Then use the fact that conjugation distributes over matrix multiplication to conjugate each of the generators, and re-assemble the resulting Pauli product.

For example, suppose we want to apply the Controlled-Y tableau to the Pauli product $X_1 Y_2$.
We start by decomposing $X_1 Y_2$ into its generators $i (X_1) (X_2) (Z_2$).
We then look up each generator in the table, replacing the generator with the result of the lookup.
This produces the result $i (X_1 Y_2) (Z_1 X_1) (Z_1 Z_2)$.
We multiply these Pauli products together to get the result $X_1$.
We conclude that $C_Y^{-1} X_1 Y_2 C_Y = X_1$.

When a Pauli product involves qubits that do not appear in the stabilizer tableau, and we want to conjugate the Pauli product inplace, those additional qubits do not need to be touched.
If the tableau covers $m$ qubits, and the Pauli product has $c$ qubits in common with the tableau, then each of the $O(c)$ table lookups has cost $O(m)$ (including multiplying together terms at the end).
Therefore, the complexity of inplace conjugation of an $n$ qubit Pauli product by an $m$ qubit stabilizer tableau with $c$ qubits in common is $O(mc)$ with no dependence on $n$ (this assumes that it is not necessary to determine which qubits are in common).
Note that $O(mc) \subseteq O(m^2)$.

In Stim, the method \path{stim::Tableau::apply_within} performs inplace conjugation of a Pauli product by a tableau.
Stim also supports out of place conjugation via the \path{stim::Tableau::operator()} operator.

\subsubsection{Composing Tableaus}

Given an $N$-qubit stabilizer tableau $A$ and an $m$-qubit stabilizer tableau $B$ where $m \leq N$, we may want to append or prepend $B$ into $A$.

Inplace appending $B$ into $A$, i.e. performing the mutation $A \rightarrow B \circ A$, is done by inplace conjugating each of $A$'s columns with $B$.
Since there are $N$ columns, and inplace conjugation by an $m$ qubit tableau takes $O(m^2)$ time, inplace appending takes $O(N m^2)$ time.

Inplace prepending $B$ into $A$ is done by computing the result of conjugating each of $B$'s generators' outputs by $A$.
For each generator $g_q$ in $B$, we compute $g_q^\prime = A^{-1} B^{-1} g_q B A$.
After computing all of the $g_q^\prime$ values, we write $g_q^\prime$ into $A$ under $g_q$ for each $g_q$ from $B$.
There are $m$ generators to conjugate by $B$ and then by $A$.
The cost of conjugating by $B$ is $O(m)$, since the generator has 1 qubit in common with $B$.
The cost of then conjugating by $A$ is $O(Nm)$, since the output from $B$ has at most $m$ qubits in common with $A$.
Therefore the complexity of inplace prepending is the same as inplace appending: $O(N m^2)$.

In Stim, inplace composition is implemented by the \path{stim::Tableau::inplace_scatter_append} method and the \path{stim::Tableau::inplace_scatter_prepend} method.
There are also many specialized optimized methods for inplace composition of common gates, such as \path{stim::Tableau::prepend_SQRT_Z}.

\subsubsection{Inverting Tableaus}

The inverse $T^{-1}$ of a stabilizer tableau $T$ is the unique tableau that satisfies $T \circ T^{-1} = T^{-1} \circ T = I$, where $I$ is the identity tableau.

Although Clifford circuits can contain highly nonlocal effects such as entanglement, inverting the Pauli terms in a stabilizer tableau is almost entirely a local process.
For example, consider the following partially specified stabilizer tableau:

$$
T = \begin{array}{r|cc|cc|cc|cc}
    & X_1 & Z_1 & X_2 & Z_2 & X_3 & Z_3 & X_4 & Z_4  \\
    \hline
    \pm & ? & ? & ? & ? & ? & ? & ? & ? \\
    1 & ? & ? & ? & ? & ? & ? & ? & ? \\
    2 & ? & ? & ? & ? & ? & ? & ? & ? \\
    3 & ? & ? & ? & ? & ? & ? & ? & ? \\
    4 & ? & ? & ? & ? & X &   & ? & ? \\
\end{array}
$$

This tableau specifies four bits of information.
It states that conjugating $X_3$ by $T$ will produce a Pauli product that has an $X$ term on qubit 4, and that conjugating $Z_3$ by $T$ produces a Pauli product with no term on qubit 4.
Nothing else is specified.

What can we determine about the inverse tableau, using only these four bits of information?
Well, we know that the generator $X_4$ commutes with $T^{-1} X_3 T$.
Therefore $T X_4 T^{-1}$ must also commute with $T T^{-1} X_3 T T^{-1} = X_3$, since conjugation preserves commutation.
This means that the term on qubit 3 of $T X_4 T^{-1}$ must be either $X$ or $I$.
Furthermore, because $X_4$ commutes with $T^{-1} Z_3 T$, we know $T X_4 T^{-1}$ must also commute with $Z_3$.
This leaves only one possibility for the term on qubit 3 resulting from conjugating $X_4$ by the inverse of $T$: the identity gate.
We can similarly solve for the term on qubit 3 of $T Z_3 T^{-1}$.
Given four bits of information about the tableau, we have derived four bits of information about the inverse tableau:

$$
T^{-1} = \begin{array}{r|cc|cc|cc|cc}
    & X_1 & Z_1 & X_2 & Z_2 & X_3 & Z_3 & X_4 & Z_4  \\
    \hline
    \pm & ? & ? & ? & ? & ? & ? & ? & ? \\
    1 & ? & ? & ? & ? & ? & ? & ? & ? \\
    2 & ? & ? & ? & ? & ? & ? & ? & ? \\
    3 & ? & ? & ? & ? & ? & ? &   & Z \\
    4 & ? & ? & ? & ? & ? & ? & ? & ? \\
\end{array}
$$

The terms on qubit $b$ that result from applying $T$ to $X_a$ and $Z_a$ always completely determine the terms on qubit $a$ from applying $T^{-1}$ to $X_b$ and $Z_b$.
To compute the Pauli terms of the inverse tableau, all that is needed is to transpose the input and output indices and then apply a few local tweaks that correspond to solving the commutation constraints.

With the Pauli terms computed, we can move on to computing the signs.
Let $S$ be the tableau with the same Pauli terms as $T^{-1}$, but with all signs positive.
If the sign in the column for the generator $g_q$ is negative, then round-tripping $g_q$ through $S$ and then $T$ will return negative $-g_q$ instead of $+g_q$.
That is to say, the sign of the $g_q$ column in $T^{-1}$ is equal to $T Sg_q S^{-1} T^{-1} g_q$.
To compute the signs, we evaluate this expression for each generator $g_q$.

Interestingly, computing the signs of the inverse tableau is more expensive than computing the Pauli terms.
It takes $O(n^2)$ time to compute the Pauli terms of the inverse of an $n$ qubit tableau, and $O(n^3)$ time to compute the signs.
One possible way to avoid this cost is to associate a pair of signs with every row of the tableau, such that the column signs of the inverse tableau are equal to the row signs of the tableau.
When appending and prepending operations into a tableau, the row signs can be updated along with the column signs at no additional cost (asymptotically speaking).
However, for the use case in this paper, only the column signs are needed so I won't bother with including the row signs.

In Stim, the inverse of a tableau can be computed using the \path{stim::Tableau::inverse} method.

\subsection{Pauli Frames}

A Pauli frame stores, for each qubit, whether or not that qubit has been bit flipped and/or phase flipped relative to some reference state.
A Pauli frame can be moved through a Clifford circuit in the same way that Pauli product observables are moved through the circuit: by conjugating the frame by the Clifford operations that the frame is passing through.
In effect, a Pauli frame is just a Pauli product where the global phase is ignored.

\subsubsection{Tracking Noise}

When simulating a noisy stabilizer circuit, where the noise is composed of probabilistic Pauli operations, it is not necessary to directly apply the noise to the simulated qubits.
Instead, the noise can be accumulated into a Pauli frame stored alongside the simulation \cite{knill2005quantum}.
As the simulation progresses, applying gates to the qubits, the Pauli frame is kept synchronized with the simulation by conjugating the frame's contents by the same gates.

When the simulation reports the measurement result from a qubit $q$, and the Pauli frame is storing an $X$ or $Y$ update for $q$, the measurement result is intercepted and inverted before being forwarded along.
The forwarded measurement are equivalent to samples from the noisy circuit.

In other words, by using a Pauli frame, you can augment any stabilizer simulator into a noisy stabilizer simulator.
This works even without access to the internal implementation details of that simulator.

\subsubsection{Tracking Corrections}

Pauli frames can also track corrective Pauli operations, so that those operations don't have to be applied to the underlying qubits.
For example, this can be used to augment a black box non-adaptive stabilizer simulator into a simulator that supports adaptive Pauli operations that depend on previous measurement results.
This even works if the simulator is replaced by a physical quantum computer \cite{ware2017experimental}.

The ability to track corrections from the outside using a Pauli frame is a key technique in quantum error correction \cite{knill2005quantum}.
It is the fundamental reason why just-in-time decoding of errors isn't needed for stabilizer codes like the surface code \cite{fowler2012surfacecodereview}.
Any corrections that are needed can be backdated into the Pauli frame at the appropriate time, and propagated forward through the circuit to the current time by toggling recorded measurements as dictated by the Pauli frame.

\section{Pauli Frame Simulation}
\label{sec:framesim}

A Pauli frame simulator works by propagating a Pauli frame through a circuit \cite{rall2019simulation}.
Clifford operations conjugate the frame, noise processes multiply Paulis into the frame, and collapsing operations randomize parts of the frame.
The benefit of Pauli frame simulation is that the frame only takes $O(n)$ bits to store and $O(1)$ time to update per gate.
The downside of Pauli frame simulation is that it doesn't tell you measurement results; it tells you whether or not measurements {\em were flipped}.
To convert this information into actual measurement results, you need a noiseless reference sample to diff against.

In a Pauli frame simulation, noise processes must be Pauli channels (i.e. equivalent to sampling a Pauli product from a probability distribution and applying the sampled product to the system's state).
For example, dephasing and depolarization are Pauli channels but relaxation to the ground state and leaking outside the computational basis aren't.
A Pauli frame simulator simulates a Pauli channel noise process by sampling a Pauli product from the Pauli product distribution corresponding to the noise process and multiplying the sampled Pauli product into the Pauli frame.

Paulis can also be multiplied into the Pauli frame by collapsing operations (initializations, resets, and measurements).
These operations introduce new stabilizers into the system.
To force later measurements to have random results if they measure observables that anticommute with a newly introduced stabilizer, the stabilizer is multiplied into the Pauli frame with 50\% probability.
In other words, after each initialization and reset and measurement, a \path{Z_ERROR(0.5)} is applied to the target qubit.
Conveniently, inserting these random Z operations is equivalent to replacing the noiseless reference sample that is being used with another independent noiseless sample from the circuit.
The random Z operations make all reference samples interchangeable and reusable.

Putting all of the pieces together, a Pauli frame simulation works as follows.
(Note: this summary will describe actions as if they were applied to one isolated Pauli frame.
Stim uses SIMD operations to apply each action to multiple frames simultaneously.)

\begin{enumerate}
  \item A reference sample is collected from the target circuit, with all noise processes disabled, using some other method.
  The reference sample can be used and reused as many times as needed.
  \item A Pauli frame is initialized with a randomly chosen $I$ or $Z$ term on each qubit (due to initialization being a collapsing operation).
  \item The Pauli frame is advanced through the circuit.
  \begin{itemize}
      \item When the frame crosses a Clifford operation $C$, the frame undergoes the update $F \rightarrow C^{-1} F C$.
      \item When the frame crosses a reset, the qubit's term in the frame is set to the identity operation.
      Then a $Z$ on the target qubit is multiplied into the frame with 50\% probability, because resets are collapsing operations.
      \item When the frame crosses a Pauli error channel, a Pauli product is sampled from the error channel (usually the no-error identity operation is by far the most likely) and multiplied into the frame.
      \item When the frame crosses a measurement, the result of that measurement (for the simulation run that the frame represents) is reported as $r_M \oplus x_q$ (where $r_M$ is the reference measurement result and $x_q$ is true if the Pauli frame has an $X$ or $Y$ term on the target qubit).
      Then a $Z$ on the target qubit is multiplied into the frame with 50\% probability, because measurements are collapsing operations.
  \end{itemize}
  \item If more samples of the circuit are needed, go to step 2. Otherwise stop.
\end{enumerate}

\section{Stabilizer Tableau Simulation}
\label{sec:tableausim}

Consider the first measurement in a stabilizer circuit.
(If there are multiple measurements tied for first, pick one of them arbitrarily.)
This measurement is measuring the observable $Z_q$, where $q$ is the qubit targeted by the measurement.
Between this measurement and the $|0\rangle$ states at start of the circuit there are a variety of Clifford gates forming a compound Clifford operation $C$.
By conjugating the current-time observable $Z_q$ by the inverse Clifford operation $C^{-1}$, we get some observable from the start of time that is equivalent to $Z_q$ at the current time.
Measuring $C Z_q C^{-1}$ at the start of time is equivalent to measuring $Z_q$ at the current time.

Suppose that $C Z_q C^{-1}$ is equal to $-Z_a Z_b$.
At the start of time we know that qubits $a$ and $b$ are in the $|0\rangle$ state, i.e. in the +1 eigenstate of the $Z$ observable.
Therefore we can replace $Z_a$ with $+1$, and we can do the same for $Z_b$.
If we were to measure $-Z_a Z_b$ at the start of time, we would get a measurement result of $-1$.
And since this observable is equivalent to $Z_q$ at the current time, we can conclude that measuring $Z_q$ now should also return a deterministic measurement result of $-1$.

Suppose alternatively that $C Z_q C^{-1}$ was equal to $-X_a X_b$.
This start-of-time  observable anticommutes with the initial state, so the measurement result will be random.
But, in order to fully resolve the measurement, we need to simplify the observable we are dealing with.
The key insight to make is that, because all qubits are initialized into the $|0\rangle$ state, a controlled operation inserted at the start of time will have no effect on the state (because the control is not satisfied).
However, inserting a controlled operation at the start of time can change the start-of-time observables.
For example, if we insert a CNOT operation controlled by qubit $a$ and targeting qubit $b$ at the start of time, we will change $C$ so that $C Z_q C^{-1}$ is equal to $-X_a$ instead of $-X_a X_b$.
We like this, because it reduces the number of terms that are present.
Measuring $Z_q$ at the current time is now equivalent to measuring $-X_a$ at the beginning of time.
The latter measurement is clearly 50/50 random and forces qubit $a$ into either the $|+\rangle$ or $|-\rangle$ state (at the start of time).
We emulate the measurement collapse by inserting a Hadamard gate at the start of time (to prepare a $|+\rangle$) and then randomly inserting an $X$ gate or not before the Hadamard (to choose between $|+\rangle$ and $|-\rangle$).
After this change to the circuit, the $Z_q$ observable at the current time is equivalent to either $Z_a$ or $-Z_a$ at the start of time, and we have reduced the random measurement case to the deterministic measurement case.

Each measurement in the circuit, working from the start of time to the end of time, can be resolved in the same way that we resolved these two example cases.
Measurements whose equivalent-start-of-time-observable contain only $Z$ terms are deterministic, and have a result equal to the sign of their equivalent-start-of-time-observable.
After reporting this result, the offending measurement gate can be deleted from the circuit.
Measurements whose equivalent-start-of-time-observable contain an $X$ or $Y$ term trigger an elimination procedure which simplifies their observable into an observable with a single $X$ term.
This can be done by introducing no-effect controlled operations and no-effect S gates at the start of time.
Once a single term remains, a Hadamard gate is inserted at the start of time to emulate the $X$ observable collapsing and an $X$ gate is randomly inserted at the start of time to emulate the randomness of the result.
This forces the measurement under consideration to become deterministic, so it can be handled by the deterministic case.

\subsection{The Asymptotic Benefits of Backwards Thinking}

At this point I should note that, in previous work \cite{aaronson2004chp}, this whole process was explained in reverse.
Instead of being framed in terms of how single qubit observables at the current time mapped to compound observables at the start of time, previous work was framed in terms of single qubit stabilizers and ``destabilizers" at the start of time being mapped to compound Pauli products at the current time.
For example, measurements were classified as deterministic or random based on whether or not they anticommuted with any of the current-time stabilizers.
This is equivalent to checking whether the $Z$ observable mapped backward anticommutes with the initial state for the same reason that inverting a Clifford tableau is nearly a transposition.

The downside of thinking in terms of mapping forwards, instead of mapping backwards, is that the measurement result is not immediately available when a measurement is deterministic.
When mapping backwards, the measurement result is the sign of the equivalent observable at the start of time.
This allows a deterministic measurement to be classified as deterministic and resolved in worst case time $O(n)$, where $n$ is the number of qubits.
When mapping forwards, it is necessary to find a combination of stabilizers which multiply together to form the desired measurement's observable.
That process has a worst case time of $O(n^2)$, instead of $O(n)$.

\subsection{Tracking the Tableau}

To perform a stabilizer tableau simulation efficiently, we need a data structure where we can efficiently append Clifford operations (as we progress further into the circuit), efficiently prepend operations (for the elimination process at the beginning of time), and efficiently determine the start-of-time observable of a measurement.
In other words, we need a stabilizer tableau.

Here are the steps used to perform a stabilizer tableau simulation:

\begin{enumerate}
    \item Initialize an identity stabilizer tableau $T$.
    To aide the intuition, imagine it being positioned at the start of the circuit being simulated, just after the qubits have been initialized into the $|0\rangle$ state.
    \item Begin folding circuit operations into $T$, working from earliest to latest.
    \begin{itemize}
        \item To fold a Clifford gate $C$ into $T$, perform an inplace prepend of $C^{-1}$ into $T$ then delete $C$ from the circuit.
        (We prepend the inverse of $C$, instead of appending $C$, because $T$ is tracking the inverse of the circuit processed so far.)
        \item To fold a measurement gate on qubit $q$ into $T$, first check whether or not the $Z_q$ column in $T$ contains any $X$ or $Y$ terms.
        If it does, the measurement is random and must be resolved.
        \begin{itemize}
            \item If the measurement is random, arbitrarily pick one of the $X$ or $Y$ terms to be the ``pivot".
            For each other $X$ or $Y$ term, perform an inplace append into $T$ of a CNOT operation controlled by the pivot targeting that term.
            (It is not necessary to remove the leftover $Z$ terms.)
            To collapse the target qubit, append into $T$ a single qubit operation targeting the pivot, changing its term to a $Z$ (from either an $X$ or $Y$).
            Finally, to randomize the measurement result, flip a coin to decide whether or not to inplace append an $X$ operation on the pivot into $T$.
        \end{itemize}
        The measurement is now deterministic.
        Report the sign of the $Z_q$ column as the result, and delete the measurement gate from the circuit.
        \item To fold a reset gate into $T$, perform a measurement on the qubit without reporting the result.
        The qubit is now either in the $|0\rangle$ state or the $|1\rangle$ state, determined by the sign of the $Z_q$ column of $T$.
        To force the qubit into the $|0\rangle$ state, overwrite the $Z_q$ column's sign with $+1$.
    \end{itemize}
\end{enumerate}

The asymptotic complexity of this simulation is $O(ng + nd + n^2r)$ where $n$ is the number of qubits, $g$ is the number of gates, $d$ is the number of measurements that have a deterministic result given previous measurements, and $r$ is the number of measurements that have random results.
(Note: $d$ and $r$ also include contributions from operations such as resets, which implicitly perform a measurement as part of their implementation.)
This is an improvement over the complexity $O(ng + n^2d + n^2r)$ from previous work.

An example of a context where this improvement provides an asymptotic advantage is simulating a distance $d$ surface code circuit for $d$ rounds.
Typically, such a circuit will have $\Theta(d^2)$ qubits, $\Theta(d^2)$ deterministic measurements in each round, and $\Theta(d^2)$ random measurements in the first and last round.
Simulating this circuit by tracking the forward tableau has a worst case simulation cost of $\Theta(d^7)$ whereas tracking the inverse tableau results in a worst case cost of $\Theta(d^6)$.
(Interestingly, the $\Theta(d^6)$ cost comes entirely the random measurements in the first and last rounds.
All of the intermediate rounds combined have a cost of $\Theta(d^5)$.)

\section{Software Engineering}
\label{sec:engineering}

\subsection{Data layout and vectorization}

There are many factors that affect software speed, such as the choice of programming language, the underlying algorithm, and company culture.
However, when attempting to write code that performs near the limits of what your computer is capable of, there are some factors that {\em have} to be considered.
Data layout is one of those factors.

In a good data layout, most data accesses are sequential.
They are performed while iterating across contiguous memory.
This makes the memory accesses predictable, so that various caching mechanisms can hide latency by prefetching data before it is needed.

In a bad data layout, data accesses are disorganized and difficult to predict, resulting in cache misses.
This can easily slow down your code by an order of magnitude (that's what it did in some quick tests I did) if not more \cite{norviglatency}.

When initially imagining Stim, I guessed that it was going to spend most of its time doing three key types of operations.
(1) Applying unitary operations and deterministic measurements to the stabilizer tableau by interacting its columns.
(2) Resolving random measurements by performing Gaussian elimination over the rows of the stabilizer tableau.
(3) Tweaking Pauli frames as they propagated through the operations in a circuit.

Notice that there are both row-wise and column-wise operations being applied to the stabilizer tableau.
This is bad, because it means we have to pick one to be fast and one to be slow.
Alternatively, we can spend effort transposing the table data to switch from one being fast to the other being fast.
In the circuits I care about (error correcting circuits) random measurements tend to come in large bursts.
With that in mind, I decided to go with the switching strategy.
Usually the stabilizer tableau is stored in column major order, so that unitary operations and deterministic measurements are efficient.
When (hopefully large batches of) random measurements need to be processed, the tableau is temporarily transposed into row major order.
(An example of a circuit that penalizes this strategy is a surface code circuit with a gauge covering a single missing data qubit \cite{nagayama2017surfacegauge}.)

I did experiment with an alternative data layout where 256x256 bit blocks from the table were contiguous in memory.
This significantly reduces the cost of switching from column-wise to row-wise operations (because only local transposes of the 256x256 blocks are needed).
However, it also forces key operations like Pauli string multiplication to skip through memory instead of iterating through contiguous memory.
Profiling showed that the costs outweighed the benefits.

For the Pauli frame simulator, I was initially worried that a good data layout didn't exist.
The problem is that the gates being performed by the circuit are not predictable, and each only affects a tiny number of bits in the Pauli frame.
This is a worst case scenario when it comes to cache misses.
Even for circuits with well organized operations, like surface code circuits operating on local 2d grids, some of the operations would have to be running ``against the grain" of memory.
I spent quite a lot of time trying to think of workarounds for this problem.

Eventually, I realized that I shouldn't be trying to vectorize within a Pauli frame simulation.
Instead, the correct thing to do is to vectorize {\em across multiple} Pauli frame simulations.
The data layout I decided on was to have 1024 frames packed together into a two dimensional table of bits, with the major axis corresponding to the qubits in the frame and the minor axis (i.e. the contiguous-in-memory axis) corresponding to the different frames being tracked.
Gates can then be applied to many frames using single instructions.
For example, an S gate targeting qubit $q$ is performed by xoring $x_q$ into $z_q$, where $x_q$ and $z_q$ are the bits of the xz-encoded Pauli for qubit $q$ in the Pauli frame.
Instead of individually operating on each $x_q$ and $z_q$ from each frame, there is a 256 bit word containing 256 separate $x_q$ bits from 256 frames, and a 256 bit word containing 256 separate $z_q$ bits from 256 frames, and the S gate is applied to all 256 frames by one \path{_mm256_xor_si256} instruction computing the 256 new $z_q$ values.
The story is similar for all other gates (e.g. see \fig{pauli_frame_code} for the CNOT gate's code).
With this approach each gate may still touch memory in an unpredictable way, but the cost of the cache miss will be amortized over a thousand parallel frame updates.

\begin{figure}
    \centering
\begin{cpp}
    int inplace_pauli_string_multiplication(
            int n, __m256i *x1, __m256i *z1 __m256i *x2, __m256i *z2) {
        // The 1s and 2s bits of 256 two-bit counters.
        __m256i c1{};
        __m256i c2{};
        // Iterate over data in 256 bit chunks.
        for (int k = 0; k < n; k++) {
            __m256i old_x1 = x1[k];
            __m256i old_z1 = z1[k];
            // Update the left hand side Paulis.
            x1[k] ^= x2[k];
            z1[k] ^= z2[k];
            // Accumulate anti-commutation counts.
            __m256i x1z2 = old_x1 & z2[k];
            __m256i anti_commutes = (x2[k] & old_z1) ^ x1z2;
            c2 ^= (c1 ^ x1[k] ^ z1[k] ^ x1z2) & anti_commutes;
            c1 ^= anti_commutes;
        }
        // Determine final anti-commutation phase tally.
        return (popcount(c1) + 2 * popcount(c2)) % 4;
    }
\end{cpp}
    \caption{
        Vectorized Pauli string multiplication.
        Multiplies one fixed length xz-encoded Pauli string into another while computing the base-$i$ logarithm of the resulting scalar phase.
        The loop body contains 4 SIMD loads, 4 SIMD stores, and 11 bitwise SIMD operations (after trivial compiler optimizations).
        I believe the number of bitwise SIMD operations is optimal, and would be very interested to hear if anyone can achieve the same effect with fewer.
    }
    \label{fig:pauli_mult_code}
\end{figure}

\begin{figure}
    \centering
\begin{cpp}
    void apply_CNOT_to_batched_pauli_frames(
            int n, __m256i *x1, __m256i *z1 __m256i *x2, __m256i *z2) {
        // Iterate over tracked frames in chunks of 256.
        for (int k = 0; k < n; k++) {
            z1[k] ^= z2[k];
            x2[k] ^= x1[k];
        }
    }
\end{cpp}
    \caption{
        Vectorized code to apply a CNOT operation to a batch of $256n$ Pauli frames.
        Note that the code is looping over multiple Pauli frames, not over multiple qubits.
        The argument \protect\path{x1} points at data storing whether or not, for each Pauli frame, there is currently an X error on the control of the CNOT (the other arguments are similar, but for the Z errors and/or the target qubit).
    }
    \label{fig:pauli_frame_code}
\end{figure}

Another data layout decision I had to make as part of implementing Stim was whether or not to interleave the x and z bits of xz-encoded Pauli products.
Initially, I decided that for locality reasons each 256 bit word should contain the 128 x bits and 128 z bits from 128 Paulis.
This is a bad idea.
It violates a core rule of thumb for writing fast vectorized code: don't mix different data together.
For example, consider simulating an $S$ gate (again).
As part of simulating the $S$ gate, the x bits of a Pauli string have to be xored pairwise into the z bits.
If the x and z bits are in separate words, you can process the xor part of the $S$ gate for 256 Paulis using one AVX bitwise xor instruction.
If the x and z bits are in the same word, the xor will have to be accompanied by some shifting and masking and these multiple instructions will have only processed 128 Paulis instead of 256.
This realization flipped my decision from ``yes interleave" to ``definitely do not interleave".
(I also  found that alternating between 256 bit x words and 256 bit z words, like \path{XZXZXZXZ}, was worse than doing all of one then all of the other, like \path{XXXXZZZZ}.
That being said, in this case the difference was small enough that I wouldn't be confident of it reproducing.)

\subsection{Sparse vs Dense}
\label{sec:sparsevdense}

Stim uses a dense representation for its stabilizer tableau.
Every bit is stored, even if almost all of the bits are zero.
In some contexts, like simulating the surface code, the stabilizer tableau representation can be made sparse and so a dense representation is wasteful.
Given how much I've been talking about the surface code in this paper, shouldn't I want to use a sparse representation instead of a dense one?
Well, maybe.
There's three issues that complicate the situation.

First, in order for a simulator to use a sparse representation, it has to {\em find} and {\em maintain} that representation.
To do this well, the simulator has to quickly solve non-local problems like ``where are the logical qubits hiding in this circuit?".
In the general case, these problems look very similar to problems with non-linear scaling (like Gaussian elimination).

Second, sparse representations hide huge constant factors compared to dense representations.
For example, Stim has a \path{--detector_hypergraph} mode where it groups error mechanisms present in a circuit based on which annotated detectors and logical observables those errors flip.
This error analysis does work very similar to what the tableau simulator does, but using a sparse representation.
Despite that, the error analysis is slower on most of the circuits I care about.
The problem is that a key bottleneck operation is computing the symmetric difference of small sets of integers, and my best efforts at writing code to do this quickly produced a result that takes on the order of ten nanoseconds to process sets with ten items.
By contrast, using a dense bit packed representation where the symmetric difference is computed using 256 bit wide xor operations, ten nanoseconds is enough time to process ten thousand items.
In other words, if you care more about speed than space, you may find that a sparse representation is worse until the sparsity of your data exceeds 99.9\%.

Third, there is elegance in a piece of software that is {\em unconditionally} fast.
It's bad enough that Stim's performance depends on the density of random measurements in a circuit.
I would prefer not to involve even more complicated properties like the existence of a stabilizer basis with low weight generators.

Using a sparse representation would be more appropriate for circuits with hundreds of thousands or millions of qubits, where even storing the stabilizer tableau becomes problematic.
Sparse representations are also more compelling in contexts where the hard or finicky parts of finding the sparse basis are embedded into the representation of the problem (e.g. explicitly annotating the locations of the logical qubits in the circuit).

\subsection{GPU Experiment Failure}

I did a small test to check if using a GPU would be beneficial for stabilizer tableau simulation or Pauli frame simulation.
All of the expensive operations in a stabilizer simulation are embarrassingly parallel bitwise operations.
They are very similar to xoring two arrays together.
So I wrote a WebGL2 fragment shader to xor together two large arrays, and tested how long it took.

I found that (surprisingly) the shader was roughly as fast at xoring as my CPU.
I expected the performance to be much better.
In hindsight, I suspect this is because the arithmetic intensity of xoring two arrays together is {\em extremely} low, and doesn't use floating point arithmetic.
That is to say, bitwise xoring doesn't hit on the comparative strengths of GPUs over CPUs.

This quick xor test suggests a GPU would perform poorly (or at least not significantly better than a CPU) at stabilizer simulation.
Since programming GPUs is harder, and was unlikely to give much benefit, I dropped the idea.

(I caution the reader that perhaps I'm simply not familiar enough with GPU programming to have produced a fast implementation.
For example, perhaps a compute shader would be more effective or perhaps I should have used a standalone GPU instead of an embedded GPU.
There's certainly no fundamental obstacle to a specialized piece of hardware running huge bitwise operations significantly faster than a CPU.)

\subsection{Threading}

I tried adding threading to various parts of Stim.
For example, when given a batch of CNOT operations to do, I tried partitioning them into two groups and running the two groups on separate threads.
Surprisingly, this was less than ten percent faster.
I didn't look deeply into why there was so little benefit; I simply decided based on the measurement that the complexity of threading batches of operations wasn't worth the gain.

One place where I found that it was worth adding threading was when transposing stabilizer tableaus.
The stabilizer tableau's data is divided into four independent pieces (x/z bits of X/Z observables), so it's trivial to process each one on a separate thread without worries of contention.
Overall I found that at large sizes this was about twice as fast as not threading the tableau transpose.

\subsection{Entropy}
\label{sec:randomness}

When running noisy Pauli frame simulations, Stim can consume gigabytes of entropy per second.
Stim gets this entropy by using \path{std::mt19937_64}, the 64 bit Mersenne twister PRNG included in C++'s standard library.
The PRNG is seeded using \path{std::random_device}, the questionable \cite{cpprandomtroubles2020} source of external entropy built into C++'s standard library.

Stim is often simulating noise that occurs with low probability (e.g. $p < 1\%$).
When deciding whether or not each instance of a noise process has occurred, Stim applies an optimization that reduces waste in the conversion from max-entropy bits to low-entropy bits.
Instead of sampling from 
\path{std::bernoulli_distribution(p)} $n$ times, Stim samples the gap between errors using \path{std::geometric_distribution(p)} $np$ times (on average).
This reduces the expected cost of sampling $n$ potential errors from $\Theta(n)$ to $\Theta(np + 1)$.

Sometimes noise occurs with intermediate probability (e.g. $10\% < p < 40\%$).
To decide whether or not this type of noise occurs, Stim uses a hybrid sampling strategy.
The desired probability $p$ is decomposed into a truncated probability $p_{\texttt{trunc}} = \lfloor 256p \rfloor / 256$ and a refinement probability $p_{\texttt{refine}} = \frac{p - p_{\texttt{trunc}}}{1 - p_{\texttt{trunc}}}$.
The truncated probability is a multiple of $2^{-8}$, so bits that are true with this probability can be sampled quickly and exactly by generating a uniformly random byte, dividing it by 256, and comparing it to the truncated probability.
Bits that are true with the refinement probability can also be sampled quickly, because the refinement probability is small.
The bitwise OR of a truncated probability bit and a refinement probability bit is a bit that is true with the desired intermediate probability.

These optimizations, and others, allow Stim to sample from any Bernoulli distribution at gigahertz rates.

\subsection{Sampling Detection Events}

A ``detector" is a specified set of measurement locations in a circuit, with the promise that xoring those locations' measurement results together will produce a deterministic value (under noiseless conditions).
When a detector's measurements xor together into a value opposite to the expected one, that is a ``detection event" indicating the presence of errors.
In addition to its measurement result sampling mode \path{--sample=#}, Stim supports a detection event sampling mode \path{--detect=#}.

Note that, in principle, detection events can be sampled with better asymptotic complexity than measurements.
For example, consider that, in the surface code, an X error on a data qubit will flip every single future measurement of a $Z$ stabilizer involving that qubit.
By contrast, when sampling detection events, this same error flips two nearby detectors and that's it.
By sampling detection events instead of measurements, errors with unbounded non-local effects can become errors with bounded local effects.
Another benefit of sampling detection events is that, since the only data that is being reported is which detectors were flipped, otherwise-distinct errors can be fused together \cite{chao2020optimization}.
Ultimately, by using a sparse representation for the set of flipped detectors, and sampling low-probability errors the way I described in \sec{randomness}, sampling the detection events in a circuit with $O(n)$ gates and detectors can be done in $O(npd + 1)$ operations (where $p$ is the average error probability and $d$ is the average number of detectors flipped by an error).
Contrast with the $O(n)$ work needed to produce measurement samples using a Pauli frame simulation.

Currently, Stim doesn't implement the asymptotically efficient detector sampling method.
Stim produces detector samples by sampling measurement results and then combining them.
This is because, for the noise levels around 0.1\% that are most interesting to me, the larger constant factors inherent in using a sparse representation outweigh the asymptotic gains.

\subsection{Testing}

Stim contains a lot of code that is very easy to get wrong.
There are hundreds of opportunities for sign errors, transposition errors, substitution errors, and omissions that would go undetected without testing.
For example, embarrassingly, version 1.0 had a transposition error where the effects of the \path{X_ERROR} and \path{Z_ERROR} noise channels were swapped when in \path{--repl} mode.

To verify the behavior of the operations in Stim, they are redundantly specified and the redundancies are checked for consistency.
Each Clifford operation is specified in at least three different ways.
As a unitary matrix:

$$
\texttt{unitary}(\sqrt{Y}) = \frac{1}{2}\begin{bmatrix}
    1 - i & 1 - i \\
    i - 1 & 1 - i
\end{bmatrix}
$$

As a stabilizer tableau:

$$
\texttt{tableau}(\sqrt{Y}) =
\begin{array}{r|cc|cc}
    & X_1 & Z_1   \\
    \hline
    \pm & + & - \\
    1 & Z & X \\
\end{array}
$$

And as hand-optimized simulator code calling vectorized subroutines:

\begin{cpp}
        void Tableau::prepend_SQRT_Y(size_t target) {
            // This method contains an intentional bug.
            xs[target].swap_with(zs[target]);
            zs[target].sign ^= 1;
        }
\end{cpp}

The unitary matrix and stabilizer tableau are validated against each other by using a vector state simulator and the state channel duality.
For each column in the tableau, the state vector simulator is initialized to contain a number of EPR pairs equal to the number of qubits that the operation under test acts on.
Let S be the subsystem made up of the first qubit from each EPR pair.
The state vector simulator applies the column's input observable to S as a Pauli product (including sign), applies the unitary matrix of the operation under test, and applies the column's output observable to S as a Pauli product (including sign).
It then verifies that the resulting state is equal (including global phase) to the state produced when acting just the unitary matrix of the operation under test on S.
This proves that the unitary matrix conjugates the input Pauli product into the output Pauli product, as specified by the tableau.

For some operations, further checks are needed before I feel confident their unitary matrix and tableau are correct.
For example, consider the $\sqrt{Y}$ operation.
The $Y$ operation has two square roots (up to global phase).
It is an arbitrary convention which of the two roots is the principle root.
What if I mixed them up?
Then I would enter the tableau for $\sqrt{Y}^\dagger$ instead of $\sqrt{Y}$, and also the unitary matrix for $\sqrt{Y}^\dagger$ instead of $\sqrt{Y}$.
Comparing the two will not catch the problem.
In these cases, I also included consistency tests which verify some disambiguating circuit identity.
For example, I want the choice of principle root to be consistent across axes, and I follow the usual convention that $\sqrt{Z} = \texttt{diag}(1, i)$.
So I can disambiguate $\sqrt{Y}$ by verifying that $\sqrt{Y} = H_{YZ} \cdot \sqrt{Z} \cdot H_{YZ} \neq \sqrt{Y}^\dagger$ where $H_{YZ} = (Y + Z) / \sqrt{2}$.

Once an operation's tableau is established as being correct, it can be used to verify the hand optimized code.
The test code does this by generating several large random tableaus \cite{bravyi2020randomtableau}, and confirming that composing the operation's tableau into the random tableau using a general method gives the same result as applying the hand optimized code.
Because of the local and discrete nature of tableau operations, this sort of fuzzing is extremely effective at catching bugs.

One important risk factor in testing is the risk of simply forgetting to test an operation in the first place.
I'd normally use code coverage tools to verify that the tests are exercising everything, but I couldn't get any C++ code coverage tool that I tried to work.
Instead, the test code takes advantage of the fact that every supported gate appears as data in a list.
This listed data is used for crucial functionality such as parsing, so I can expect it to be complete.
With this listed data, one test can verify all of the defined gates.
For example:

\begin{cpp}
    TEST(gate_data, tableau_data_vs_unitary_data) {
        for (const auto &gate : GATE_DATA.gates()) {
            if (gate.flags & GATE_IS_UNITARY) {
                EXPECT_TRUE(tableau_agrees_with_unitary(
                    gate.tableau(), gate.unitary())) << gate.name;
            }
        }
    }
\end{cpp}

Noise channels are harder to test than Cliffords, because the behavior of noise channels isn't deterministic.
I tested noise channels by creating tests that sampled circuits applying the noise channel to simple states.
Then, statistical checks verified the consistency of the samples against independent definitions of noise channels from Cirq \cite{quantum_ai_team_and_collaborators_2020_4062499}.
The \path{X_ERROR}/\path{Z_ERROR} transposition that I mentioned slipped through these tests because originally they only exercised the bulk sampling API (which was a natural choice due to needing many samples), not the single-sample interactive API.

Tests are run with compiler optimizations enabled, but also run without compilation optimizations with address and memory sanitization checks enabled.
A continuous integration system verifies that the tests pass before allowing commits to be merged into the main branch.

\subsection{Profiling and Optimizing}

To determine the performance of individual components of Stim, I wrote a small benchmarking framework that would repeatedly run some task and compare the time it took to a reference time.
For example, here is code benchmarking the multiplication of Pauli strings with ten thousand terms:

\begin{cpp}
    BENCHMARK(PauliString_multiplication_10K) {
        size_t n = 10 * 1000;
        PauliString p1(n);
        PauliString p2(n);
        benchmark_go([&]() {
            p1.ref().inplace_right_mul_returning_log_i_scalar(p2);
        }).goal_nanos(90).show_rate("Paulis", n);
    }
\end{cpp}

Note that an important assumption here is that the compiler is inlining the \path{benchmark_go} method and the lambda.
I found that making \path{benchmark_go} a template with the lambda's type as a parameter caused more consistent inlining, but I can make no guarantees.

\begin{figure}
    \centering
    \resizebox{0.99\linewidth}{!}{
        \includegraphics{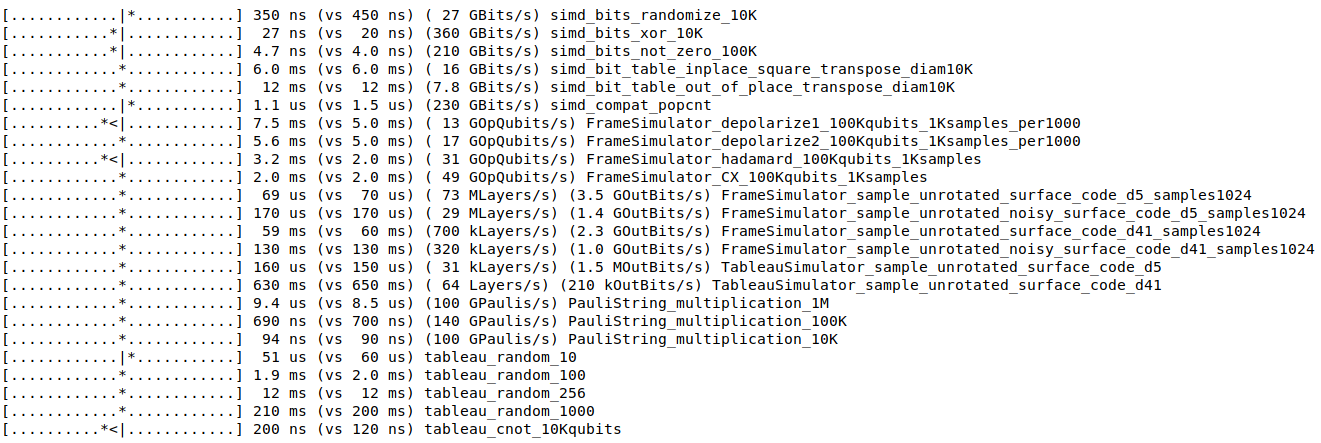}
    }
    \caption{
        Screenshot of results from \texttt{stim\_benchmark}.
        Each line is the result of a small benchmark defined in Stim's source code.
        The ASCII bar indicators on the left show deviations from the reference time in a visual way that can easily be scanned over.
        The location of the asterisk, relative to the center of the ASCII bar, indicates how many decibels (factors of $1.26$) slower or faster the run was (compared to the reference time).
        In the screenshot, a few benchmarks are running slow.
        That could indicate some sort of performance regression, or could just be noise.
    }
    \label{fig:stim-benchmark-output}
\end{figure}

The framework runs whatever is in the \path{benchmark_go} body for 0.5 seconds, counting how many iterations were finished.
It then produces output comparing the inferred time to a reference time.
Example benchmarking output can be seen in \fig{stim-benchmark-output}.
This benchmarking methodology would not work well for a team of people (e.g. because different computers require different reference times), but was sufficient for my purposes.

When I wanted to improve the performance of a benchmark, or fix a regression, I would build Stim using debug flags \texttt{-g -fno-omit-framepointer} and run it under Linux's \path{perf} tool to identify functions and lines of code taking disproportionate amounts of time.
For example, the reason that I optimized the Bernoulli distribution sampling process to use less entropy per error is because profiling showed that generating entropy for errors was a significant bottleneck.

\subsection{Portability}

One of the major downsides of using AVX instructions is that they are not portable.
They won't work on older CPUs, and they won't work on architectures besides x64 (e.g. most phones).
Stim tries to work around this by hiding all AVX instructions behind an opaque type \path{simd_word}.
Stim includes one implementation of \path{simd_word} that uses 256-bit-wide AVX instruction, another that uses older 128-bit-wide SSE instructions, and another that uses no non-standard instructions.
The presence of defines like \path{__AVX2__} then determine which \path{simd_word} implementation is used.

Because the size of a \path{simd_word} is not fixed, code is written in a fashion that avoids depending on this size.
For example, the class \path{simd_bits}, which represents a bit packed array that supports vectorized instructions, will help the caller with this by automatically padding its size up to a multiple of \path{simd_word}'s size.
\path{simd_bits} also exposes methods which handle the boilerplate parts of iterating over the words.
\path{simd_bit_table} plays a similar role for two dimensional data.

\subsection{Input Format}

Users need some way of describing quantum circuits to Stim.
I considered using an established format for this, such as OpenQASM.
I dropped the idea because OpenQASM has a lot of semantics that would be time consuming to implement (such as functions, named variables, and file includes).
Also OpenQASM is currently undergoing a breaking version change \cite{openqasmbreakingchange}.

Consequently, I decided to make yet another input format.
Stim's input format is nearly minimal.
It takes a series of lines, and each line can specify a gate to apply to some targets.
For example, the line ``\texttt{H 0}" says to apply a Hadamard to qubit 0 and the line ``\texttt{CNOT 0 1}" says to apply a Controlled-NOT operation controlled by qubit 0 targeting qubit 1.
That's essentially all that's necessary to know in order to use the format, besides the names of supported gates.

There are of course several exceptions to this minimalist ideal.
Blank lines are permitted.
Lines can be suffixed with a comment prefixed by a hash ``\texttt{\# like this}".
Gates can be broadcast over multiple targets like ``\texttt{H 0 1 2}".
And, most shamefully, there is a ``\texttt{REPEAT N \{...\}}" macro which repeats a block of instructions a given number of times (drastically reducing some file sizes).

\subsection{Output Format}

By default, Stim outputs a series of `0' and `1' characters, one for each measurement result.
The final measurement result is followed by a newline character `\textbackslash n'.
When multiple shots are taken, the measurement results (and newline) from the first shot are output, then the results (and newline) from the second shot, and so forth.

Stim also supports a binary output format, triggered by providing the command line argument \path{--out_format=b8}.
This format packs 8 measurements results into each byte of output, ordered from least significant bit to most significant bit.
The last byte output for a shot is padded to completion with 0s.
There is no separator between the bytes from separate shots.

In addition to these ``dense" formats, Stim supports sparse output formats like \path{--out_format=hits} and \path{--out_format=r8} that focus on the indices of non-zero bits.
These output formats are useful when sampling detectors, because non-zero bits are rarer.

\section{Comparison}
\label{sec:compare}

I compared Stim's performance to Aaronson and Gottesman's \path{chp} simulator \cite{aaronson2004chp}, IBM's Qiskit simulator (\texttt{qiskit method=`stabilizer'}) \cite{Qiskit}, Google's Cirq simulator (\path{cirq.CliffordSimulator}) \cite{quantum_ai_team_and_collaborators_2020_4062499}, and finally Anders and Briegel's \path{graphsim} \cite{anders2006fastgraphsim} simulator.

I performed the comparison benchmarks on my work laptop: a ThinkPad running gLinux (a Google internal modified variant of Debian) with an Intel Core i7-8650U CPU @ 1.90GHz and 16GB of RAM.
I compiled \path{stim} and \path{chp} with \texttt{g++ v10.2.1} using optimization level \path{-O3}.
I generated circuit files understood by each simulator (samples of these files are included in the ancillary files of this paper), and then iterated through the files of each benchmark timing how long each simulator took.
Output from the simulators was discarded, not recorded.
I didn't do anything special to prepare the computer for benchmarking except ensure it was plugged in with a full battery with no other user programs running.
I didn't disable dynamic frequency scaling or auto-updating mechanisms (the simulators differ in performance by large enough factors that I don't think this is a problem).

Timing was done by calling \path{time.monotonic()} or \path{std::chrono::steady_clock::now()} before and after the key simulation methods and computing the difference (I edited source code to do this when necessary).
Timing data doesn't include program startup or circuit parsing, except for ``\texttt{stim with startup+parsing}" where I used \texttt{date +\%s\%N} before and after invoking Stim in Bash.

Because the various simulators I benchmarked don't consume a common format or support a common gate set, it was often necessary to decompose one gate into several for one of the simulators.
For example, \path{chp} doesn't have a Z gate so two S gates were used instead when needed.
Additionally, because \path{graphsim} is an API rather than an end-to-end tool, I had to write a bit of glue code to drive that API using data read from a file.
For simplicity, I wrote this glue code to use \path{chp}'s file format which has very few gates.

After running the benchmarks I noticed that there were outlier times at the smallest sizes.
I think this is because the last problem size in each benchmark was large, and so simulating it could for example consume a lot of memory and evict pages that would then fault during the next run (which would be the smallest case in the next benchmark).
To correct this, I discarded the suspicious data for the three smallest cases from each benchmark and re-ran them on their own.

\begin{figure}
    \centering
    \resizebox{\linewidth}{!}{
        \includegraphics{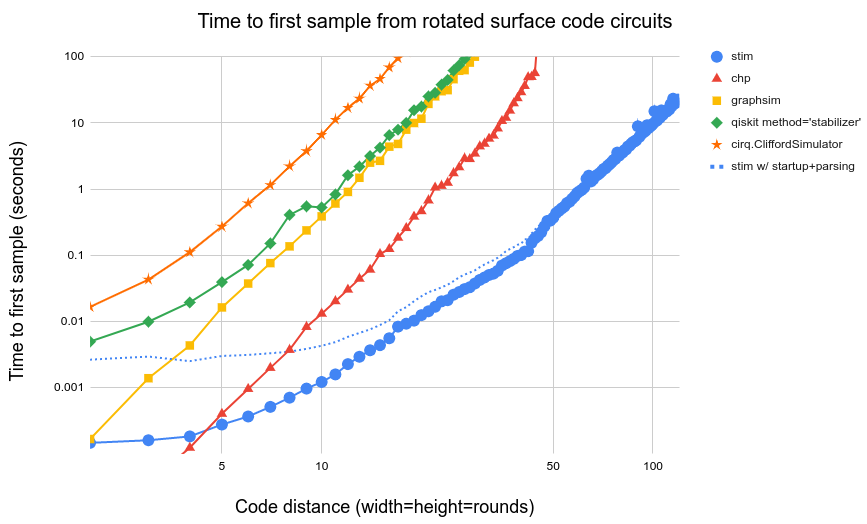}
    }
    \caption{
        Delay until first sample on $d \times d \times d$ unrotated surface code memory circuits \cite{horsman2012latticesurgery}.
        The intention of this benchmark is to highlight simulators that can take advantage of redundant structure in the surface code, such as the abundance of local stabilizers and deterministic measurements.
        Based on the apparent scaling, it appears that none of the simulators are making much use of the structure of the surface code.
        Stim scales comparatively well on this benchmark because Stim performs deterministic measurements in worst case linear time instead of quadratic time.
    }
    \label{fig:bench-surface}
\end{figure}

\begin{figure}
    \centering
    \resizebox{\linewidth}{!}{
        \includegraphics{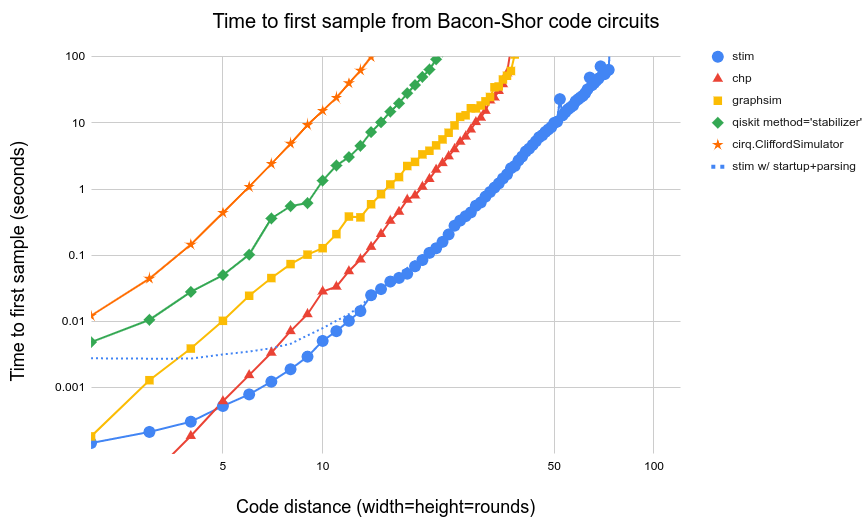}
    }
    \caption{
        Delay until first sample on $d \times d \times d$ Bacon-Shor code memory circuits \cite{bacon2006operator}.
        The intention of this benchmark is to contrast with the surface code benchmark, in that the Bacon-Shor code also has a lot of structure but doesn't have lots of deterministic measurements.
        Note how Stim's apparent scaling on this benchmark looks more like the other simulators than it did in the surface code benchmark.
    }
    \label{fig:bench-bacon}
\end{figure}

\begin{figure}
    \centering
    \resizebox{\linewidth}{!}{
        \includegraphics{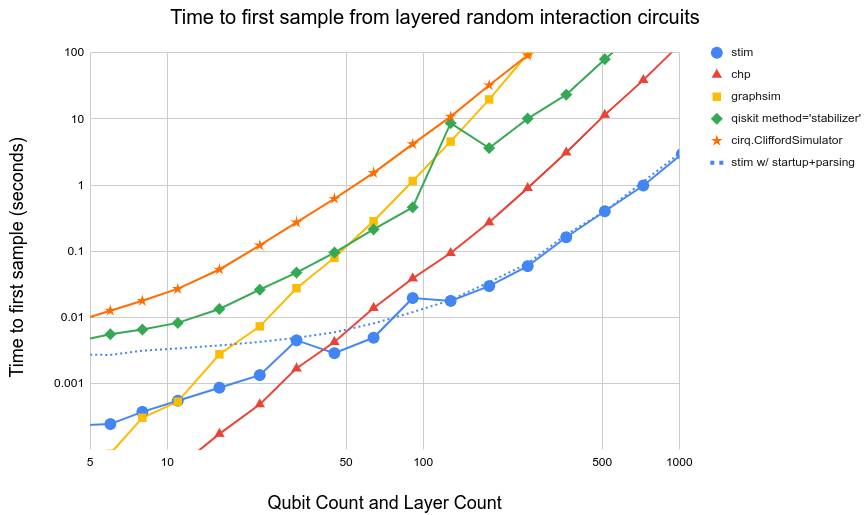}
    }
    \caption{
        Delay until first sample on random circuits.
        The intention of this benchmark is to highlight simulators with good performance in circuits that lack much exploitable structure.
        Each circuit is made up of $n$ qubits operated on for $n$ layers.
        Each layer randomly applies an $H$, $S$, or $I$ gate to each qubit, then samples a random pairing of the qubits and applies a CNOT to each pair, then samples 5\% of the qubits to measure in a random basis.
        At the end of the circuit, every qubit is measured in a random basis.
        One effect I don't understand here is why Stim appears to be scaling differently from chp.
    }
    \label{fig:bench-random}
\end{figure}

\begin{figure}
    \centering
    \resizebox{\linewidth}{!}{
        \includegraphics{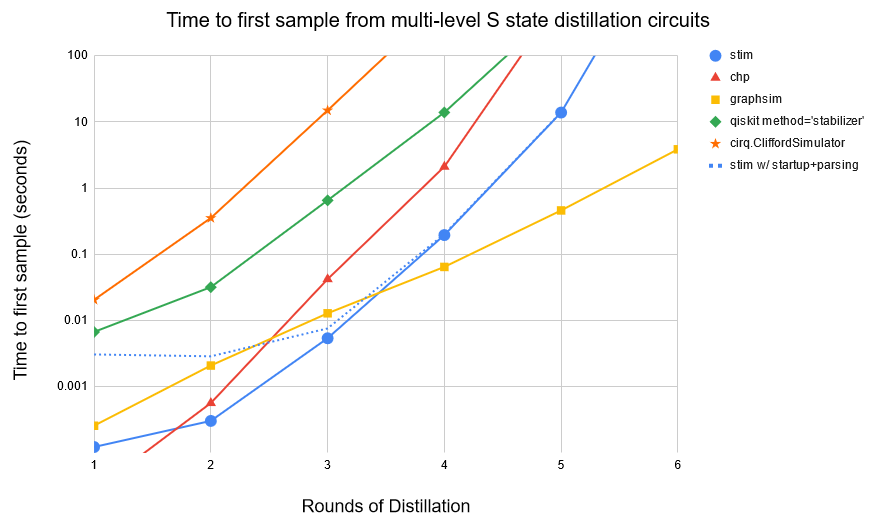}
    }
    \caption{
        Delay until first sample on nested 7-to-1 S state distillation circuits \cite{fowler2012surfacecodereview}.
        The intention of this benchmark is to highlight simulators that do well on circuits made up of nearly independent pieces.
        For example, with 5 levels of distillation, there are initially 2401 independent pieces (each with 15 qubits) performing identical copies of the 7-to-1 distillation circuit \href{https://algassert.com/quirk\#circuit=\%7B\%22cols\%22\%3A\%5B\%5B1\%2C1\%2C1\%2C1\%2C1\%2C1\%2C1\%2C1\%2C1\%2C1\%2C\%22\%E2\%80\%A2\%22\%2C\%22X\%22\%2C\%22X\%22\%2C\%22X\%22\%5D\%2C\%5B1\%2C1\%2C1\%2C1\%2C1\%2C1\%2C1\%2C1\%2C1\%2C\%22\%E2\%80\%A2\%22\%2C1\%2C\%22X\%22\%2C1\%2C\%22X\%22\%2C\%22X\%22\%5D\%2C\%5B1\%2C1\%2C1\%2C1\%2C1\%2C1\%2C1\%2C1\%2C\%22\%E2\%80\%A2\%22\%2C1\%2C1\%2C\%22X\%22\%2C\%22X\%22\%2C1\%2C\%22X\%22\%5D\%2C\%5B\%22\%E2\%80\%A2\%22\%2C1\%2C1\%2C1\%2C1\%2C1\%2C1\%2C1\%2C1\%2C1\%2C1\%2C1\%2C\%22X\%22\%2C\%22X\%22\%2C\%22X\%22\%5D\%2C\%5B1\%2C\%22\%5E\%3DA7\%22\%2C1\%2C1\%2C1\%2C1\%2C1\%2C1\%2C\%22inputA7\%22\%5D\%2C\%5B1\%2C1\%2C1\%2C1\%2C1\%2C1\%2C1\%2C1\%2C\%22H\%22\%2C\%22H\%22\%2C\%22H\%22\%2C\%22H\%22\%2C\%22H\%22\%2C\%22H\%22\%2C\%22H\%22\%5D\%2C\%5B1\%2C\%22Measure\%22\%2C\%22Measure\%22\%2C\%22Measure\%22\%2C\%22Measure\%22\%2C\%22Measure\%22\%2C\%22Measure\%22\%2C\%22Measure\%22\%2C\%22Measure\%22\%2C\%22Measure\%22\%2C\%22Measure\%22\%2C\%22Measure\%22\%2C\%22Measure\%22\%2C\%22Measure\%22\%2C\%22Measure\%22\%5D\%2C\%5B1\%2C\%22inputA7\%22\%2C1\%2C1\%2C1\%2C1\%2C1\%2C1\%2C\%22\%5E\%3DA7\%22\%5D\%2C\%5B1\%2C1\%2C1\%2C1\%2C1\%2C1\%2C1\%2C1\%2C\%22X\%22\%2C\%22X\%22\%2C\%22X\%22\%2C\%22\%E2\%80\%A2\%22\%5D\%2C\%5B\%22Z\%22\%2C1\%2C1\%2C1\%2C1\%2C1\%2C1\%2C1\%2C\%22X\%22\%2C1\%2C\%22X\%22\%2C1\%2C\%22\%E2\%80\%A2\%22\%5D\%2C\%5B\%22Z\%22\%2C1\%2C1\%2C1\%2C1\%2C1\%2C1\%2C1\%2C1\%2C\%22X\%22\%2C\%22X\%22\%2C1\%2C1\%2C\%22\%E2\%80\%A2\%22\%5D\%2C\%5B\%22Z\%22\%2C1\%2C1\%2C1\%2C1\%2C1\%2C1\%2C1\%2C\%22X\%22\%2C\%22X\%22\%2C1\%2C1\%2C1\%2C1\%2C\%22\%E2\%80\%A2\%22\%5D\%2C\%5B1\%2C1\%2C1\%2C1\%2C1\%2C1\%2C1\%2C1\%2C\%22Chance3\%22\%5D\%2C\%5B\%22Bloch\%22\%2C1\%2C1\%2C1\%2C1\%2C1\%2C1\%2C1\%2C\%22\%7C0\%E2\%9F\%A9\%E2\%9F\%A80\%7C\%22\%2C\%22\%7C0\%E2\%9F\%A9\%E2\%9F\%A80\%7C\%22\%2C\%22\%7C0\%E2\%9F\%A9\%E2\%9F\%A80\%7C\%22\%5D\%5D\%2C\%22init\%22\%3A\%5B\%22-\%22\%2C\%22i\%22\%2C\%22i\%22\%2C\%22i\%22\%2C\%22i\%22\%2C\%22i\%22\%2C\%22i\%22\%2C\%22i\%22\%2C\%22\%2B\%22\%2C\%22\%2B\%22\%2C\%22\%2B\%22\%5D\%7D}{(view in Quirk)}.
        Then the output qubits from each of these pieces are put into groups of seven and handed along to the next stage, which has 343 independent pieces.
        This process continues until there is a single output qubit remaining.
        graphsim does particularly well on this benchmark, due to the connected components of the graph state perfectly matching the independent pieces of the circuit.
    }
    \label{fig:bench-distill}
\end{figure}

I chose five benchmarking tasks: bulk sampling a surface code circuit (see \fig{bench-surface-1000}), sampling a surface code circuit (see \fig{bench-surface}), sampling a Bacon-Shor code circuit (see \fig{bench-bacon}), sampling a randomly generated circuit (see \fig{bench-random}), and sampling multi-level S state distillation (see \fig{bench-distill}).
See the figure captions for commentary on each benchmark task.

Overall, I would say that the benchmarks show that at larger sizes Stim outperforms everything else by one or more orders of magnitude, and that when collecting thousands of samples Stim outperforms everything else by many orders of magnitude.
A notable exception is the multi-level distillation task, where \path{graphsim} shines due to its ability to simulate piecewise circuits in a piecewise fashion.
Actually, in most of the benchmarks, there is circuit structure present that should allow a simulator that notices and uses that structure to outperform Stim.
This would be an interesting avenue for future work.

\section{Conclusion}
\label{sec:conclusion}

In this paper I presented Stim, a fast stabilizer circuit simulator.
Stim was my attempt at simulating surface code circuits with tens of thousands of qubits and millions of operations in a tenth of a second.
Although that goal wasn't reached, Stim is regardless a useful piece of software.

I still think it's possible to simulate stabilizer circuits with tens of thousands of qubits and millions of gates in a tenth of a second, at least in special cases.
For example, when a person analyzes a surface code circuit, they notice the repetitiveness of the low level details of the circuit and zoom out to focus on topological details like lattice surgeries and braids.
If the stabilizer simulator could do something similar, decent speedups should be possible.
That is to say, one avenue of attack is to improve the performance of finding and maintaining sparse representations.

Another avenue of attack is to re-examine the assumptions going into the problem being solved.
The motivating use case I described at the start of the paper was a user making small iterative changes to a large circuit.
In that context the typical simulation is not a brand new circuit, but rather a slight variation of a circuit from just a moment ago.
This suggests that work could be cached and reused.
Alternatively, perhaps as part of creating the circuit, the user has provided information that can be used by the simulator.
For example, the user may have specified the circuit in terms of repeated pieces or may have noted that a measurement is supposed to relate to previous measurements in a specific way.
The simulator could verify this information (instead of having to find it) and then exploit it.

Ultimately, I hope the experimental and theoretical sides of the quantum community find Stim to be a useful tool.
When you can do something ten or a hundred times faster, new and interesting use cases become possible.
For example, with a fast enough simulator, you could consider brute forcing (or machine learning) the internal minutia of an error correction circuit to minimize undesirable error propagation properties.
I certainly intend to take advantage of the speed in my own work.

\section{Acknowledgements}

I thank Michaels Broughton and Newman for putting up with excessive jokes about {\em brute speed}.
I thank Michael Broughton for advice and discussions on profiling and optimizing code.
I thank N. Cody Jones for useful discussions about simulation algorithms, and in particular for pointing out that randomizing the Z frame after resets and measurements was sufficient for sampling from the space of noiseless circuit outputs.
I thank Hartmut Neven for creating an environment where this work was possible in the first place.

\bibliographystyle{plainnat}
\bibliography{refs}

\appendix

\section{Short Usage Examples}
\label{app:example}

\subsection{(Command Line) Sample a noisy repetition code 1024 times}

\begin{python}
    echo -e "
        REPEAT 20 {\n
            X_ERROR(0.01) 0 1 2 3 4 5 6 7 8 9 10\n
            CNOT 0 1 2 3 4 5 6 7 8 9\n
            CNOT 10 9 8 7 6 5 4 3 2 1\n
            MR 1 3 5 7 9\n
        }
    " | ./stim --sample=1024
\end{python}

\subsection{(Python) Sample a noisy repetition code 1024 times}

\begin{python}
    import stim
    distance = 5
    rounds = 20
    bit_flip_rate_per_round = 0.01
    qubits = range(2 * distance + 1)
    # Build circuit.
    c = stim.Circuit()
    c.append_operation("X_ERROR", qubits, bit_flip_rate_per_round)
    c.append_operation("CNOT", qubits[:-1])
    c.append_operation("CNOT", qubits[::-1][:-1])
    c.append_operation("MR", qubits[1::2])
    c *= rounds
    # Collect and print samples.
    sampler = c.compile_sampler()
    samples = sampler.sample(1024)
    for sample in samples: print(sample)
\end{python}

\subsection{(Python) Interactive quantum teleportation}

\begin{python}
    import stim
    s = stim.TableauSimulator()
    # Share EPR pair.
    s.h(1)
    s.cnot(1, 9)
    # Prepare state to teleport.
    s.h(0)
    s.s(0)
    # Alice measurements.
    s.cnot(0, 1)
    s.h(0)
    x, z = s.measure_many(1, 0)
    # Bob corrections.
    if x: s.x(9)
    if z: s.z(9)
    # Check received correctly by uncomputing preparation.
    s.s_dag(9)
    s.h(9)
    should_be_false = s.measure(9)
    print(f"Alice got {x}, {z}. Bob uncomputed to {should_be_false}")
    assert should_be_false == False
\end{python}

\subsection{(Python) Algebraic manipulations}

\begin{python}
    import stim

    t = stim.Tableau.random(10)
    t_inverse = t**-1
    assert t * t_inverse == stim.Tableau(10)
    assert (t**1000000) * (t_inverse**500000)**2 == stim.Tableau(10)

    x5: stim.PauliString = t.x_output(5)
    z6: stim.PauliString = t.z_output(6)
    assert t_inverse(x5 * z6) == stim.PauliString("+_____XZ___")
    
    s = stim.TableauSimulator()
    s.set_inverse_tableau(t)
    stabilizers = s.canonical_stabilizers()
    assert stabilizers[0].commutes(stabilizers[1])
\end{python}

\end{document}